\documentclass[aps,pra,amssymb,twocolumn,superscriptaddress,10pt,tightenlines]{revtex4-2}

\usepackage[english]{babel}
\usepackage{dcolumn}
\usepackage{bm}
\usepackage{amsfonts}
\usepackage{amsmath}
\usepackage{amssymb,amsthm}
\usepackage{mathtools}
\usepackage{braket}
\usepackage{physics}
\usepackage{color}

\usepackage{subfigure}
\usepackage{url}
\usepackage{epsfig}
\usepackage{hyperref}
\usepackage{graphicx}
\usepackage{booktabs}
\usepackage{makecell}
\usepackage{multirow}
\usepackage{comment}
\usepackage{enumerate}

\usepackage[normalem]{ulem}

\newcommand{\sinc}{\operatorname{sinc}}

\begin{document}

\title{Condensation of vanishing photon emission rates in random atomic clouds \\
}
\author{Viviana Viggiano}
\affiliation{Dipartimento di Fisica, Universit\`a di Bari, I-70126 Bari, Italy}
\affiliation{INFN, Sezione di Bari, I-70125 Bari, Italy}
\author{Romain Bachelard}
\affiliation{Departamento de F\'isica, Universidade Federal de S\~ao Carlos, Rodovia Washington Lu\'is, km 235-SP-310, 13565-905 S\~ao Carlos, S\~ao Paulo, Brazil}
\author{Fabio Deelan Cunden}
\affiliation{Dipartimento di Matematica, Universit\`a di Bari, I-70125 Bari, Italy}
\affiliation{INFN, Sezione di Bari, I-70125 Bari, Italy}
\author{Paolo Facchi}
\affiliation{Dipartimento di Fisica, Universit\`a di Bari, I-70126 Bari, Italy}
\affiliation{INFN, Sezione di Bari, I-70125 Bari, Italy}
\author{Robin Kaiser}
\affiliation{Universit\'e C\^ote d'Azur, CNRS, Institut de Physique de Nice, 06200 Nice, France}
\author{Saverio Pascazio}
\affiliation{Dipartimento di Fisica, Universit\`a di Bari, I-70126 Bari, Italy}
\affiliation{INFN, Sezione di Bari, I-70125 Bari, Italy}
\author{Francesco V. Pepe}
\affiliation{Dipartimento di Fisica, Universit\`a di Bari, I-70126 Bari, Italy}
\affiliation{INFN, Sezione di Bari, I-70125 Bari, Italy}
\author{Antonello Scardicchio}
\affiliation{The Abdus Salam International Center for Theoretical Physics, Strada Costiera 11, 34151 Trieste, Italy}
\affiliation{INFN Sezione di Trieste, Via Valerio 2, 34127 Trieste, Italy}
%\date{May 28, 2025}

\begin{abstract}
In the collective photon emission from atomic clouds both the atomic transition frequency and the decay rate are modified compared to a single isolated atom, leading to the effects of superradiance and subradiance. In this article, we analyse the properties of the Euclidean random matrix associated to the radiative dynamics of a cold atomic cloud, previously investigated in the contexts of photon localization and Dicke super- and subradiance. We present evidence of a new type of phase transition, surprisingly controlled by the cooperativeness parameter, rather than the spatial density or the diagonal disorder. The numerical results corroborate the occurrence of such a phase transition at a critical value of the cooperativeness parameter, above which the lower edge of the spectrum vanishes exhibiting a macroscopic accumulation of eigenvalues. Independent evaluations based on the two phenomena provide the same value of the critical cooperativeness parameter.
\end{abstract}

\maketitle

\noindent

\section{Introduction} 

Many-body physics and emerging collective effects present many surprising features with theoretical, numerical and experimental challenges. A fruitful approach to discover novel features often exploits interdisciplinary research and concepts, not directly focusing on detailed implementation and applications. Motivated by the recent development of open quantum many-body systems, in this article we investigate novel features based on a detailed analysis of the eigenvalues of a random matrix involved in the description of collective decay processes. 
Following Dicke's pioneering article~\cite{Dicke}, the investigation of cooperative effects in atom-photon interactions has garnered increasing attention, both from a theoretical and experimental point of view. The interaction of photons with atomic ensembles produces complex collective phenomena~\cite{review_coop}, which profoundly differ from the behaviour of individual components of the system. For instance, in the collective photon emission from atomic clouds~\cite{molmer} both the atomic transition frequency~\cite{lamb,lamb2} and the decay rate are modified compared to a single isolated atom. 
This rate can be larger or smaller than the decay rate $\Gamma$ of an isolated atom, yielding \textit{superradiance}~\cite{exp_sup,essay,intro,direct2}  and \textit{subradiance}~\cite{b0,exp_sub,sub_kaiser2021, sub2_kaiser2021} respectively. The possibility to control the emergence of collective phenomena paves the way for a wide range of applications, e.g.\ for quantum information processing and quantum communication~\cite{interfaces,quantum_internet, appl_super, appl_sub}. 
Motivated by this ongoing effort on cooperative effects and light localisation in two-level systems, implemented for instance in cold atomic clouds, we focus on the spectral properties of a Euclidean random matrix, relevant for photon localization and Dicke super- and subradiance~\cite{Svidzinsky2008, Svidzinsky2010, Skipetrov2011}. The same random matrix was the subject of previous analyses on photon localisation~\cite{Akkermans08} and cooperativeness~\cite{Viggiano2023} (see \cite{vectorial} for a vector model generalization). 
We note that phase transitions in different Euclidean random matrices have been identified with a phase transition either controlled by the spatial density of the scatterers~\cite{Skipetrov14, Bellando14} or by additional diagonal disorder~\cite{Celardo24}. 
In this article we show that a new type of phase transition has been found in the Euclidean random matrix as studied in~\cite{Akkermans08, vectorial, Viggiano2023}, surprisingly controlled by the so-called cooperativeness parameter, different from the spatial density and the diagonal disorder parameter. Even though the approach we follow is based on eigenvalue statistics rather than observables of a driven system or an evolution of an interacting many-body system, understanding an eigenvalue problem allows for a complementary insight in the underlying emerging properties of complex systems.

\section{The physical system and the matrix $S$} 

The system under investigation is based on a random distribution of resonant scatterers, such as in a cold cloud of $N$ identical atoms with fixed random positions $\bm{r}_j$, $j=1,\ldots,N$, independently sampled from a three-dimensional Gaussian distribution with zero mean and variance $\sigma^2>0$~\cite{exp_sub,exp_sup, K,Viggiano2023}.  
We treat the atoms as two-level systems with transition frequency $\omega_a=c k_a$.
For the atom-photon interaction we use a dipole model in the rotating wave and scalar approximations~\cite{Cohen1,Cohen} (the vectorial case has been addressed, e.g.\ in~\cite{vectorial}). 
We do not consider near-field dipole-dipole coupling and thus resort to the so-called scalar model. \textit{A posteriori}, this will be justified by the low-density nature of the limit~\cite{Viggiano2023}.
Furthermore, we work in the weak excitation limit, restricting ourselves to the
single excitation sector, with at most one emitted photon or one atom in the excited state at a time.

\begin{figure*}[ht]
\includegraphics[width=\textwidth]{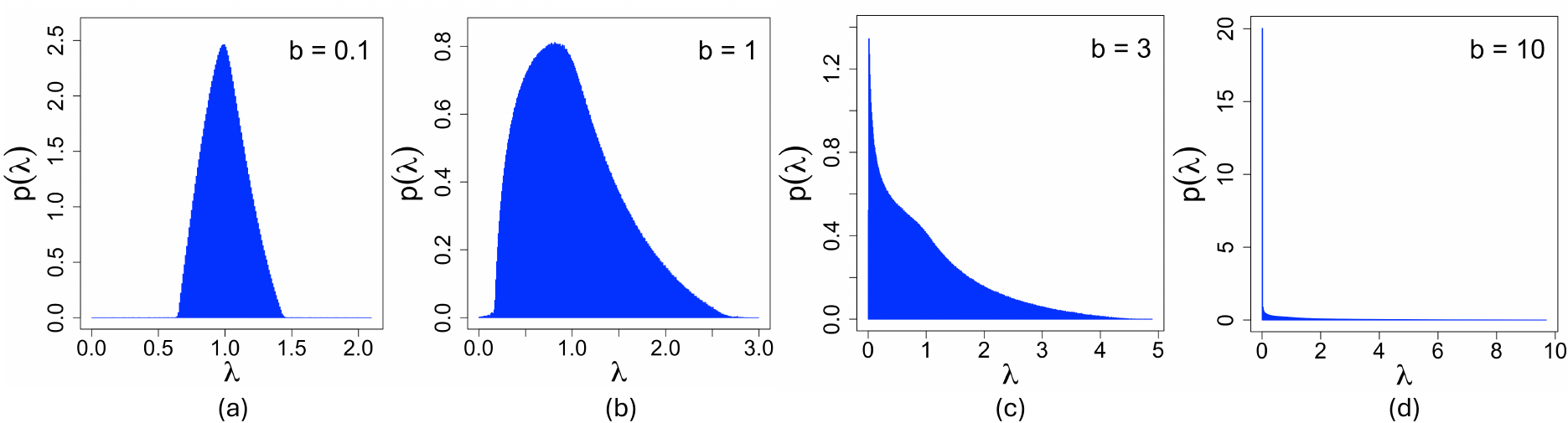}
\caption{Histograms of the normalized eigenvalue distribution $p(\lambda)$ obtained from 10 realizations of the matrix $S$ in Eq.~\eqref{eq:Sb0} with $N=20000$ and different values of the cooperativeness parameter $b$.}
\label{fig:spettro}
\end{figure*}

In this setting, let us denote with $\beta_j(t)$ the probability amplitude that the $j$th atom
is excited. The total excitation probability of the atomic cloud is $P(t)=\bm{\beta}^{\dagger}(t)\bm{\beta}(t)$, where $\bm{\beta}(t)$ is a column vector with components $\beta_j(t)$.
After switching off the laser, the evolution equation of the excitation probability reads~\cite{exp_sup, K, Courteille}
\begin{equation} 
\dot{P}(t)=- \Gamma \bm{\beta}^\dagger(t)S\bm{\beta}(t)
    \label{eq:Pev2}\,,
\end{equation}
where $\Gamma$ is the linewidth of the atomic transition. The emission rate matrix $S$ is a Euclidean random matrix~\cite{Bogomolny,Parisi,Goetschy,Bordenave}, whose elements are given by
\begin{equation}
S_{ij}=  \sinc\left(\sqrt{\frac{N}{b }}\|{\vb*{x}_i - \vb*{x}_j}\| \right), \quad i,j=1,\ldots, N,
 \label{eq:Sb0}
\end {equation}
with $\bm{x}_i=\bm{r}_i/\sigma$ the rescaled dimensionless atomic positions. These are independently sampled from a three-dimensional standard Gaussian probability density function
\begin{equation}
p(\vb*{x})=\frac{1}{(2\pi)^{3/2}} e^{-\|{\vb*{x}}\|^2/2},\quad \vb*{x} \in \mathbb{R}^3, 
    \label{eq:gaussian}
\end{equation}
for which the \textit{cooperativeness parameter} $b$ is defined as
\begin{equation}
 b =\frac{N}{(k_a \sigma)^2}
    \label{eq:b0}\,.
\end{equation}
Note that this parameter is proportional to the optical depth used in other works, but differs by a numerical factor. While photon-mediated interactions also induce coherent dynamics \cite{Viggiano2023,AsenjoGarcia2017}, diagonalizing the emission rate matrix \eqref{eq:Sb0} provides direct information on independent decay channels, since its elements represent, in the Weisskopf-Wigner approximation, the overlap between the asymptotic states $\int d\bm{k} e^{-i\bm{k}\cdot\bm{r}_j}(ck- ck_a +i\Gamma/2)^{-1}\ket{\bm{k}}$ of photons emitted by atoms in $\bm{r}_j$ \cite{Cohen}.

It was shown that the cooperativeness parameter $b$ is the relevant scaling parameter both from a physical and a mathematical point of view to describe the cooperative nature of the cloud emission, see Refs.~\cite{exp_sub,exp_sup,b0,CD,Viggiano2023}.
Henceforth, we turn to studying the asymptotic behavior of $S$ in Eq.~\eqref{eq:Sb0} when both the numerator and the denominator in~\eqref{eq:b0} tend to infinity, with their ratio
kept fixed~\cite{Viggiano2023}. More specifically, we investigate the spectrum of $S$ in the limit $N\to\infty$ for various finite values of $b$. Notice that this corresponds to a vanishing atomic density and finite optical depth limit. Indeed the dimensionless peak density, i.e.\ the density at the center of the Gaussian cloud multiplied by the cube of the wavelength $\lambda_a=2\pi/k_a$, is given by
$\rho_0 \lambda_a^3=(2\pi b )^{3/2} N^{-1/2}$
and hence vanishes in the limit $N\to\infty$ with $b$ fixed. Though a vector model of photon-mediated interaction would generally provide a more detailed description, we show in Appendix~\ref{app:vector} that the scalar and vector models provide equivalent information in the limit of $N\to\infty$ and fixed cooperativeness parameter, up to a rescaling of the latter \cite{vectorial}. Therefore, the results obtained for the investigated phase transition can be generalized to the vector model, while the differences between models become crucial for Anderson localization \cite{Skipetrov14}, emerging at finite density.

The matrix $S$ is positive semidefinite and satisfies the trace condition $\Tr S=N$. Therefore, the eigenvalues $\lambda_i$,  $i=1,\ldots,N$ of $S$ are nonnegative,   satisfy $\sum_i\lambda_i = N$, and have unit mean. See~\cite{Viggiano2023} for more details.
The eigenvalue $\lambda=1$ corresponds to the decay rate $\Gamma$ of a single isolated atom; $0 \leq \lambda<1$ corresponds to the subradiant modes of the matrix $S$ in~\eqref{eq:Sb0}, with a longer lifetime;  $1<\lambda \leq N$ is associated with superradiant modes, with a decay rate larger than $\Gamma$. Figure~\ref{fig:spettro} displays the  eigenvalue distribution $p(\lambda)$ for several values of the cooperativeness parameter $b$, with a distribution very different for each value considered. For small values of $b$ the spectrum is symmetric around the isolated-atom emission peak at $\lambda=1$, while for increasing $b$ the effects of super- and subradiance become more pronounced. For large values of $b$ [Fig.~\ref{fig:spettro}(d)] the spectrum is characterized by an accumulation of eigenvalues in the subradiant sector, compensated by the presence of very few large eigenvalues.

\begin{figure*}
\centering
\includegraphics[width=\textwidth]{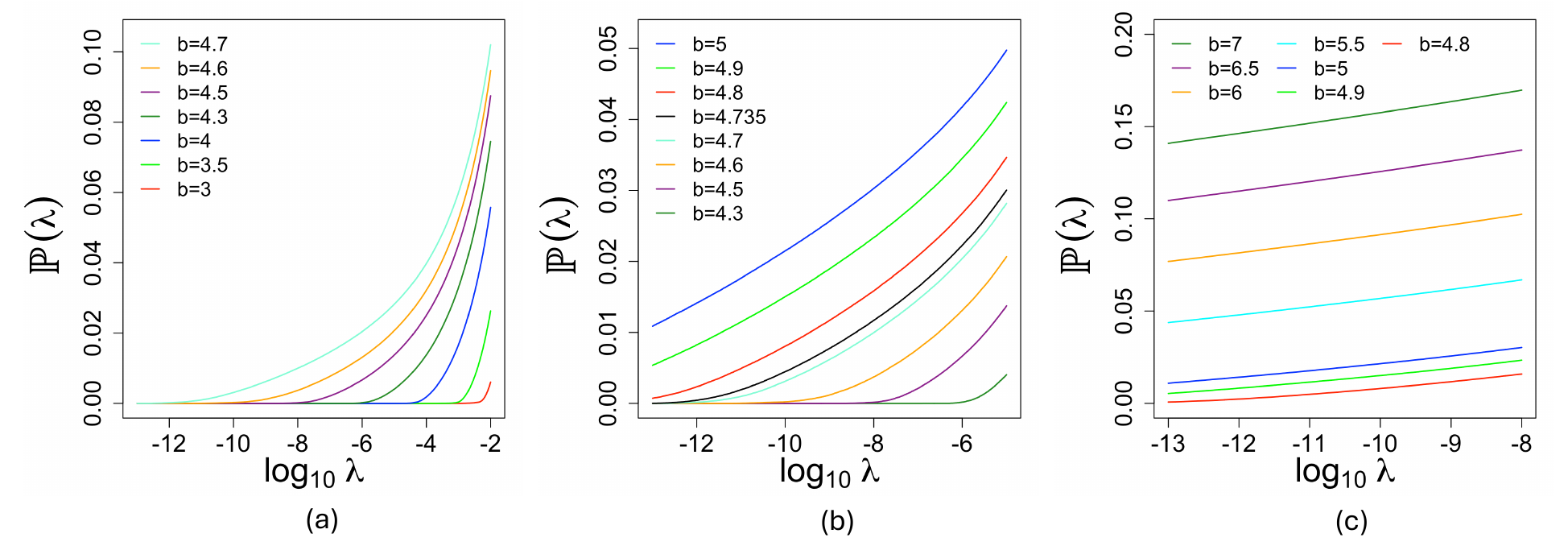}
\caption{Cumulative distribution $\mathbb{P}(\lambda)=\int_0^\lambda p(y) \mathrm{d} y$ of small eigenvalues for (a) $b <b_{c}$; (b) $b$ close to $b_{c}$; (c) $b >b_{c}$. The distributions are obtained with $N=20000$.}
\label{fig:ccc}
\end{figure*}

In this article we focus on the very subradiant part of the spectrum, when $\lambda \ll 1$. 
It is manifest in Fig.~\ref{fig:spettro} that the spectrum displays a significant accumulation of small eigenvalues as $b$ is increased. These eigenvalues signal the presence of an important subradiant component in the dissipative dynamics of the atomic gas. We want to investigate this feature and in particular scrutinize the phenomenon of the accumulation of vanishing eigenvalues. We shall see that this corresponds to a phase transition which is controlled by the cooperativeness parameter $b$.

Our strategy is the following. We first analyze the behaviour of the minimum eigenvalue $\lambda_{\min}$ and prove that it vanishes above a threshold value $b_c$ of the cooperativeness and in the $N\to\infty$ limit. Then we 
compute the limit of the fraction of vanishing eigenvalues $N_0/N$. We anticipate that this corresponds to a finite fraction of eigenvalues vanishing in the $N\to\infty$ limit, which appears at the value $b=b_c$ at which the density of eigenvalues touches the value $\lambda=0$. All these features point to the presence of a phase transition.

\section{Small eigenvalues distribution}

Let us first analyze the cumulative distribution 
\begin{equation}
 \mathbb{P}(\lambda)=\int_0^\lambda p(y) \mathrm{d} y   
\end{equation} 
of the emission rates 
for small $\lambda$'s ($\lambda<10^{-2}$) for different values of the cooperativeness $b$. The cumulative distribution is displayed in Fig.~\ref{fig:ccc} for different values of $b$ in the range $[3,7]$. The numerical precision of our simulations is $10^{-13}$, which means that eigenvalues below $10^{-13}$ are, in practice, indistinguishable from 0. 
We notice the absence of eigenvalues in the interval between $\lambda=0$ and $\lambda_{\min}>0$ in Fig.~\ref{fig:ccc}(a) for $b\leq 4.7$, which becomes smaller as $b$ tends to a critical value $b_{c}\approx 4.7$.  In Fig.~\ref{fig:ccc}(b) the cumulative distributions in the range $10^{-13} \leq \lambda \leq  10^{-5}$, for $4.3 \leq b  \leq 5$, are monitored, and we observe that the interval shrinks to zero at 
\begin{equation}
b_{c} = 4.73(5).
 \label{eq:bc}
\end{equation}
At this point, a single non-decaying mode is born, with infinite lifetime (in the $N\to\infty$ limit). As the cooperativeness $b$ is increased above $b_c$, we also observe the appearance of the fast rising of $\mathbb{P}(0)$. As one can observe in Fig.~\ref{fig:ccc}(c), for $4.8 \leq b\leq 7$, this behavior is compatible with a {\it finite fraction} of vanishing eigenvalues and therefore non-decaying modes, which increases as $b$ is increased beyond $b_c$. The asymptotic extent of this fraction will be scrutinized in what follows.

\subsection{Study of the minimum eigenvalue} 

\begin{figure}
    \centering
    \includegraphics[width=0.75\columnwidth]{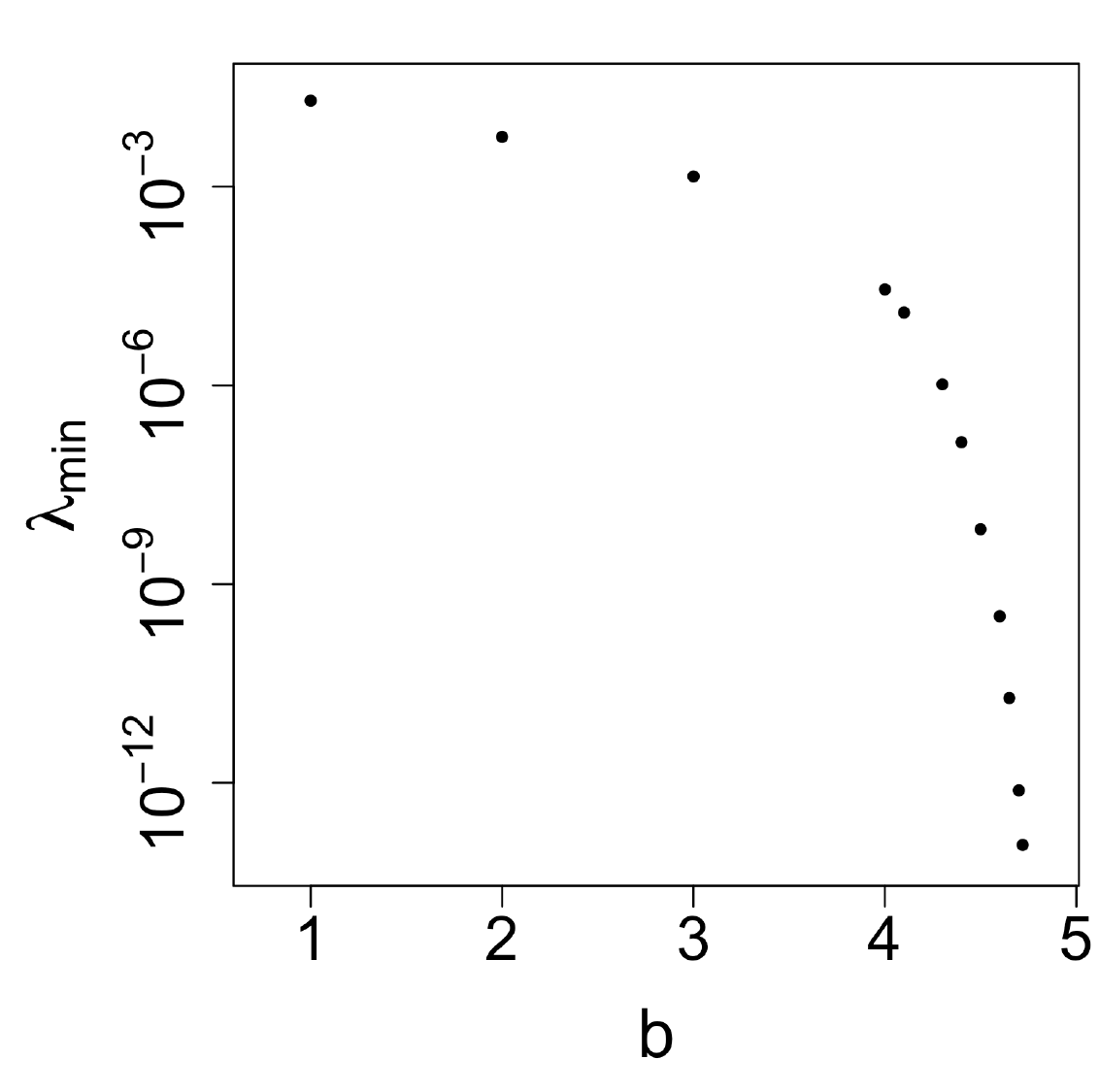}
    \caption{Plot of the minimum emission rate $\lambda_{\mathrm{min}}$ as a function of $b$ below the critical value $b_c=4.735$. Notice the log scale on the vertical axis. Here $\lambda_{\mathrm{min}}$ is the minimum eigenvalue obtained for the corresponding $b$ and for $N=30000$, which is the largest achievable matrix size in numerical simulations.}\label{fig:lambdamin_b0}
\end{figure}

Let us now focus on the minimum eigenvalue of the spectrum of $S$, corresponding to the most subradiant state. We have taken into account several values of $b$ and computed the mean minimum eigenvalue $\lambda_{\min}$, averaged over several realizations of the matrix $S$, for different values of atom number $N$. 
An evidence of the transition at $b_{c}$ emerges from Fig.~\ref{fig:lambdamin_b0}, in which $\lambda_{\mathrm{min}}$ is plotted as a function of $b$. Notice that here $\lambda_{\mathrm{min}}$ is the minimum eigenvalue obtained for the corresponding $b$ and for the largest achievable matrix 
For all the values $b>4.735$, $\lambda_{\min}$ falls below our numerical precision, thus being compatible with zero. Also, $\lambda_{\min}$ displays a scaling which is fully compatible with 0 in the $N\to\infty$ limit, see Appendix~\ref{app:minimum}.

The presence of the transition at $b_{c} = 4.735$ can also be deduced from Fig.~\ref{fig:lambdamin_b0}, by the first derivative of the curve $\lambda_{\mathrm{min}}$ vs $b$: It has a tangent compatible with the vertical in correspondence of the critical point (notice the log scale on the vertical axis). 

We conclude that the behavior of the minimum eigenvalue represents a first quantitative evidence of a phase transition at $b_c$. Indeed, phase transitions related to the the eigenvalue distribution touching the origin have been observed in other physical systems, e.g.\ in the entanglement spectrum of a bipartite system~\cite{matrix08}.

The occurrence of a phase transition associated to the sudden divergence of $p(0)$, driven by increasing $b$, is made evident by the study of the convergence of a locator expansion perturbation theory \cite{forward1,qrem,infdloc,RRG1,RRG2}. This argument, presented in Appendix~\ref{app:estimate}, captures the origin of the transition as well as the correct scaling with $N$, predicting a critical value that overestimates the numerical result by a factor $\sim 2$. However, what that argument cannot explain is the fact that the minimum eigenvalue $\lambda_{\min}$ touches $0$ exactly at the same time at which a finite density of eigenvalue is formed at $\lambda=0$, as shown below. Controlling $\lambda_{\min}$ requires a degree of control on the large deviations of the density $p(\lambda)$, which typically is not captured by perturbative calculations. We are therefore led to rely on the numerical results, which however seem quite robust.

\begin{figure}
    \includegraphics[width=0.5\textwidth]{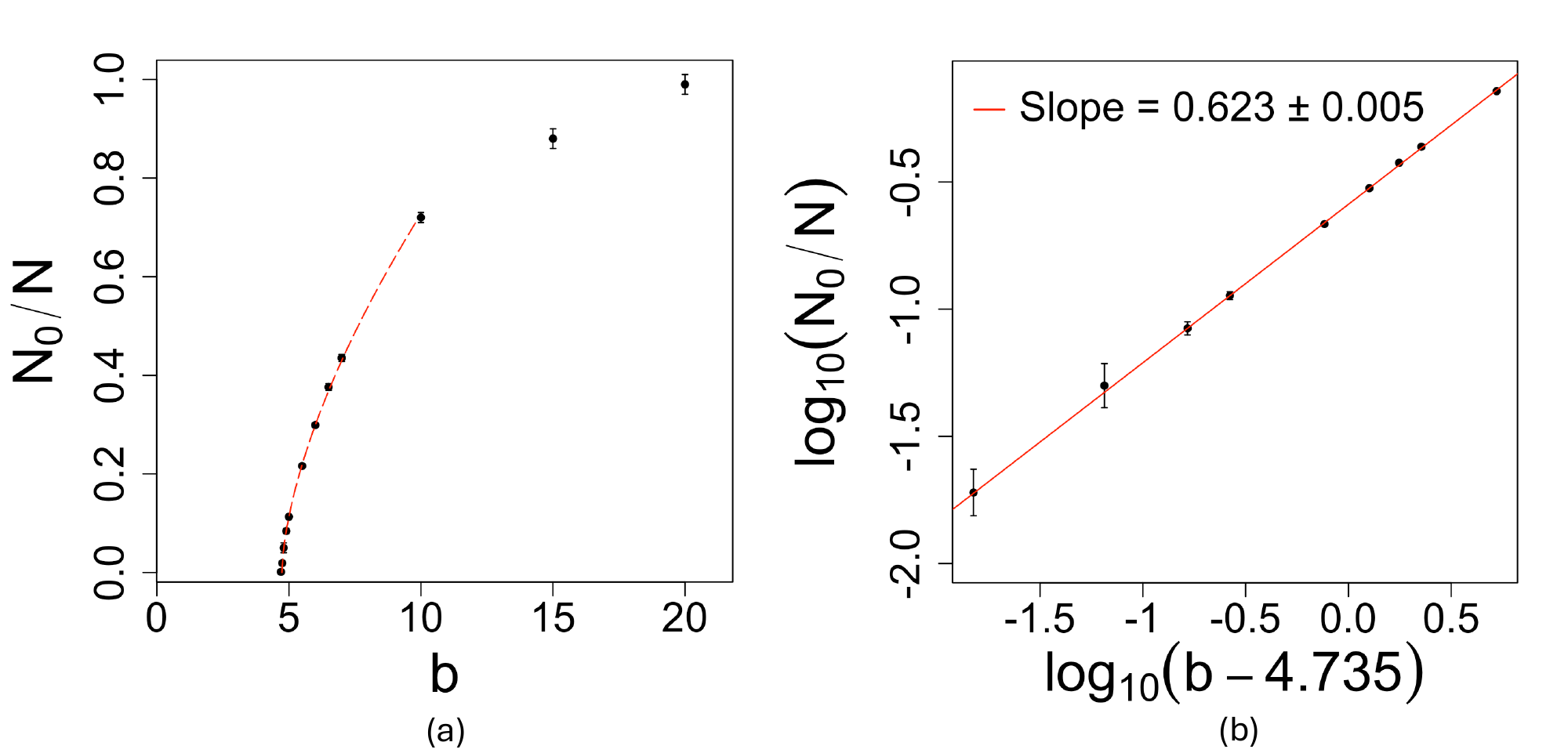}
    \caption{Asymptotic fraction of vanishing eigenvalues as a function of $b$. Panel (a) shows that for $b <4.7$ there are no eigenvalues below $10^{-13}$, for $b =4.9$ the condensate fraction is less than $0.1$ and it increases as $b$ grows, approaching unity for $b =20$.  Panel (b) shows the linear relationship between $\log_{10}(N_0/N)$ and $\log_{10}(b -b_{c}')$, where $b'_c=4.735$ is the critical value above which the spectrum exhibits an accumulation of vanishing eigenvalues. The slope of the linear fit yields the critical exponent $\beta\simeq0.623$ of the power-law~\eqref{eq:powerlaw}. This power-law has been used to fit the transition region in Panel (a) (red dashed curve). Notice that the critical vale $b'_c$ coincides with the value $b_c$ determined by the study of the minimum eigenvalue.}\label{fig:picco_vs_b0}
\end{figure}

\subsection{Fraction of eigenvalues in the condensate}

For $b  \gtrsim b_c$ the spectrum of $S$ starts to exhibit an accumulation of  eigenvalues close to zero, which becomes larger for increasing values of $b$. We count the fraction of these  eigenvalues for different values of $b$ and $N$. Considering the finite numerical precision, we count as part of this \emph{condensate} all the eigenvalues of $S$ smaller than $10^{-13}$.

At fixed $b$, we count the fraction  $N_0/N$ of vanishing eigenvalues and, as explained in Appendix~\ref{app:condensate}, extrapolate for $N \to \infty$ the fraction of the condensate at 0. 
The evolution of this fraction as a function of $b$ is presented in Fig.~\ref{fig:picco_vs_b0}. In panel (a), the presence of a transition at a critical value close to $4.7$ can be observed. Below this threshold there is no vanishing eigenvalue (even for large matrix sizes), while above the threshold the asymptotic fraction of vanishing eigenvalues grows. It approaches unity for $b =20$, indicating that almost all eigenvalues vanish in this regime.

It is interesting to estimate the critical value $b_{c}'$ above which the spectrum exhibits the transition toward an accumulation of vanishing eigenvalues, and to assess whether near this critical point the fraction of subradiant vanishing eigenvalues grows according to a power law:
\begin{equation}
\frac{N_0}{N} \propto (b -b_{c}')^{\beta}\,,
    \label{eq:powerlaw}
\end{equation}
where $\beta$ is a critical exponent. 
To this end, we focus on the data in Fig.~\ref{fig:picco_vs_b0}(a) corresponding to values of $b$ from 4.75 to 10. 
These data are used to plot $\log_{10}(N_0/N)$ as a function of $\log_{10}(b -b_{c}')$ for several values of $b_{c}'$ close to $4.7$, and for each of them we perform a linear fit -- thus checking whether Eq.~\eqref{eq:powerlaw} is satisfied. From this protocol, we estimate the critical value $b_{c}'$ as the one yielding the best linear regression fit.

The result of this fitting procedure is shown in Fig.~\ref{fig:picco_vs_b0}(b), for the optimal value of $b_{c}'$, indicating that indeed the fraction of vanishing eigenvalues obeys the power-law~\eqref{eq:powerlaw} near the transition, with a critical exponent $\beta\simeq 0.623$. Notice that the critical value $b_{c}'$ is compatible with the value $b_c$ in Eq.~\eqref{eq:bc}, found for the birth of the non-decaying eigenmode, thus demonstrating the consistence between the different approaches to characterize the transition. The power-law~\eqref{eq:powerlaw} has been used to fit the transition region in Fig.~\ref{fig:picco_vs_b0}(a) (red dashed curve). We note that the occurrence of weak phase transitions with a resulting formation of a nonzero singular component at the edge of the eigenvalue spectrum is a universal phenomenon which has been reported in many physics problems related to random matrices~\cite{third18,thirdlong19} and in random networks \cite{Metz20,Silva22}. 

We finally remark that the Markovian assumption of neglecting photon propagation delays is well motivated for optical atomic transitions (see \cite{exp_sub}), where non-Markovian effects cannot emerge at distances smaller than tens of meters.

\section{Conclusions}

The results obtained point to the existence of a phase transition, occurring at a critical value $b_{c}$ of the cooperativeness parameter $b$. Increasing $b$ beyond $b_c$, the smallest eigenvalue $\lambda_{\mathrm{min}}$ of the matrix $S$ approaches the extremely-subradiant boundary $\lambda= 0$ and the spectrum starts exhibiting an accumulation of vanishing eigenvalues in the subradiant sector. In other words, there is a single transition point, which we find to be at $b_{c}=4.73(5)$ at which both phenomena occur: For $b  \geq b_{c}$ the smallest eigenvalue in the asymptotic limit $N \to +\infty$ reaches zero and the spectrum begins to display an accumulation of vanishing eigenvalues. We identify a mechanism for which the presence of eigenvalues at $0$ can occur by analyzing the perturbation theory in small $b$ for the density of eigenvalues $p(0)$, showing that it diverges above a certain $b_{c}^{0}$. Our perturbative value of $b_c^0=11.2$ {\it overestimates} the numerically obtained $b_c=4.73(5)$, in line with our understanding of the validity of this kind of derivation.

We note that while the Anderson localization transition has been predicted for a fixed atomic density~\cite{Skipetrov14, Bellando14}, the transition reported here rather occurs, in the large-$N$ limit, at a vanishing atomic density, but at a finite optical depth. In this sense, it is directly related to the cooperativeness of the system \eqref{eq:b0}, which is the control parameter for super- and subradiance~\cite{review_coop}. In the dilute limit, vanishing two-body interactions are compensated by both their long-range nature and the growing number of coupled atoms, providing another signature of the cooperative nature of the transition. In this context, it would be particularly interesting to understand how disorder on the energy coupling, either from the dipole-dipole excitation-exchange (Hamiltonian) terms or from energy broadening (diagonal disorder) could affect the emission rate condensation transition. Finally, we note that in an experimental setup, the vanishing emission rate transition may compete with decoherence mechanisms other than spontaneous emission, such as finite-temperature effects~\cite{Weiss2019,motional_effects2020,motional_effects2023}, which may set a lower bound on the emission rates~\cite{Lassegues}.

\begin{acknowledgments}
\textit{Acknowledgments.---}
This work is supported by Regione Puglia and \mbox{QuantERA} ERA-NET Cofund in Quantum Technologies (GA No.\ 731473)
and the French National Research Agency (ANR19-QUAN-003-01), project PACE-IN.
FDC, PF and SP acknowledge the support from Istituto Nazionale di Fisica Nucleare (INFN) through the project `QUANTUM'. FDC and PF acknowledge the support from the Italian National Group of Mathematical Physics (GNFM-INdAM). FDC acknowledges the supports from PRIN 2022 project 2022TEB52W-PE1-
‘The charm of integrability: from nonlinear waves to random matrices’.
PF acknowledges the supports from PNRR MUR project CN00000013 `Italian National Centre on HPC, Big Data and Quantum Computing'.
AS acknowledges funding by the European Union - NextGenerationEU under the project NRRP “National Centre for HPC, Big Data and Quantum Computing (HPC)'' CN00000013 (CUP D43C22001240001) [MUR Decree n. 341- 15/03/2022] - Cascade Call launched by SPOKE 10 POLIMI: “CQEB” project.
SP and FVP acknowledge the support from PNRR MUR project PE0000023-NQSTI.
R.K. acknowledges support from the European project ANDLICA, ERC Advanced Grant Agreement No. 832219.
R.K. and R.B. acknowledge funding from ANR-FAPESP (projectQuaCor ANR19-CE47-0014-01) and project STIC-AmSud (Ph879-17/CAPES 88887.521971/2020-00) and CAPES-COFECUB (CAPES 88887.711967/2022-00)
\@. RB has received the financial support of the S\~ao Paulo Research Foundation (FAPESP) (Grants No.~2018/15554-5, 2019/13143-0, 2022/00209-6 and 2023/03300-7).
\end{acknowledgments}

\appendix

\section{Connection with the vector model}
\label{app:vector}

Consider the imaginary part of the vector Green's matrix that describes the pairwise interaction between dipoles in positions $\{\bm{r}_i\}_{i\in\{1,\dots,N\}}$ \cite{vectorial}
\begin{equation}\label{eq:Lambda}
    \Lambda_{i\alpha,j\beta} = S_1(k_a r_{ij}) P_{\perp,\alpha\beta} (\hat{\bm{r}}_{ij}) + S_2(k_a r_{ij}) P_{\parallel,\alpha\beta} (\hat{\bm{r}}_{ij}) ,
\end{equation}
with $\bm{r}_{ij}=\bm{r}_i-\bm{r}_j$, $r_{ij}=|\bm{r}_{ij}|$ and $\hat{\bm{r}}_{ij} = \bm{r}_{ij}/r_{ij}$, where the distance-dependent functions are defined as
\begin{align}
    S_1 (x) & = \frac{3}{2} \left( \frac{\sin(x)}{x} + \frac{ x \cos(x) - \sin(x) }{x^3} \right) , \\
    S_2 (x) & = -3\, \frac{ x \cos(x) - \sin(x) }{x^3} ,
\end{align}
and
\begin{equation}
    P_{\perp,\alpha\beta} (\hat{\bm{r}}) = \delta_{\alpha\beta} - \hat{r}_{\alpha}\hat{r}_{\beta}, \qquad P_{\parallel,\alpha\beta} (\hat{\bm{r}}) = \hat{r}_{\alpha}\hat{r}_{\beta}
\end{equation}
are the transverse and longitudinal projectors, respectively. The terms $\Lambda_{i\alpha,i\beta}=\delta_{\alpha\beta}$ are consistent with the $x\to 0$ limits of $S_1$ and $S_2$. This immediately provides the result
\begin{equation}
    \mathrm{Tr}(\Lambda) = 3N .
\end{equation}
The computation of
\begin{align}
    \mathrm{Tr}(\Lambda^2) & = \sum_{i,j=1}^N \sum_{\alpha,\beta=x,y,z} \left( \Lambda_{i\alpha,j\beta} \right)^2 \nonumber \\ 
    & = 3N + \sum_{i\neq j} \sum_{\alpha,\beta=x,y,z} \left( \Lambda_{i\alpha,j\beta} \right)^2
\end{align}
is simplified by considering that
\begin{align}
    & \sum_{\alpha\beta} \left( P_{\perp,\alpha\beta} (\hat{\bm{r}} ) \right)^2 = \sum_{\alpha} P_{\perp,\alpha\alpha} (\hat{\bm{r}}) = 2 , \\
    & \sum_{\alpha\beta} \left( P_{\parallel,\alpha\beta} (\hat{\bm{r}} ) \right)^2 = \sum_{\alpha} P_{\parallel,\alpha\alpha} (\hat{\bm{r}}) = 1 , \\
    & \sum_{\beta} P_{\perp,\alpha\beta} (\hat{\bm{r}}) P_{\parallel,\alpha\beta} (\hat{\bm{r}}) = 0 .
\end{align}
Therefore,
\begin{equation}
    \mathrm{Tr}(\Lambda^2) = 3N + \sum_{i\neq j} \left( 2 S_1^2 (k_a r_{ij}) + S_2^2(k_a r_{ij}) \right) .
\end{equation}
If the dipole positions are random, with independent and identical distributions, the expectation value of $\mathrm{Tr}(\Lambda^2)$ can be used to estimate the second moment of the distribution of the eigenvalues $\lambda$ of $\Lambda$:
\begin{equation}
    \left\langle \lambda^2 \right\rangle = \frac{\left\langle \mathrm{Tr}(\Lambda^2) \right\rangle}{3N} = 1 + \frac{N-1}{3} \left( 2 \left\langle S_1^2 (k_a r) \right\rangle + \left\langle S_2^2 (k_a r) \right\rangle \right) .
\end{equation}
In the case of an isotropic Gaussian distribution with standard deviation $\sigma$ along the three directions, the dimensionless distances $R=r/\sigma$ are distributed according to the probability density \cite{Viggiano2023}
\begin{equation}
    p(R) = \frac{1}{\sqrt{4\pi}} R^2 \exp \left( - \frac{R^2}{4} \right) .
\end{equation}
Defining the dimensionless parameter $M=(k_a\sigma)^2$, we obtain
\begin{widetext}
\begin{align}
    \left\langle S_1^2 ( k_a r ) \right\rangle = \int_0^{\infty} dR \, S_1^2 (\sqrt{M} R) p(R) = & \, \frac{3}{64 M^3} \Bigl[ 12 M^2 + 6 M + 1 - 8 \sqrt{\pi} M^{3/2} \mathrm{erf}(2\sqrt{M}) - e^{-4M} ( 12 M^2 + 10 M + 1 ) \Bigr] , \\
    \left\langle S_2^2 ( k_a r ) \right\rangle = \int_0^{\infty} dR \, S_2^2 (\sqrt{M} R) p(R) = & \, \frac{3}{16 M^3} \Bigl[ 1 - 6 M + 4 \sqrt{\pi} M^{3/2} \mathrm{erf}(2\sqrt{M}) + e^{-4M} ( 2 M - 1 ) \Bigr] .
\end{align}
\end{widetext}
As in the scalar case, the variance $(\Delta\lambda)^2 = \left\langle \lambda^2 \right\rangle - \left\langle \lambda \right\rangle^2 = \left\langle \lambda^2 \right\rangle - 1$ converges to a finite value in the $N\to\infty$ limit only if $M$ scales linearly with $N$. In this case, defining 
\begin{equation}
    b = \frac{N}{M} = \frac{N}{(k_a\sigma)^2} ,
\end{equation}
we obtain
\begin{equation}
    \left\langle \lambda^2 \right\rangle \to 1 + \frac{3}{8} b ,
\end{equation}
with the only nontrivial contribution coming from $\left\langle S_1^2 ( k_a r ) \right\rangle$.

\begin{figure}
    \centering
    \includegraphics[width=0.35\textwidth]{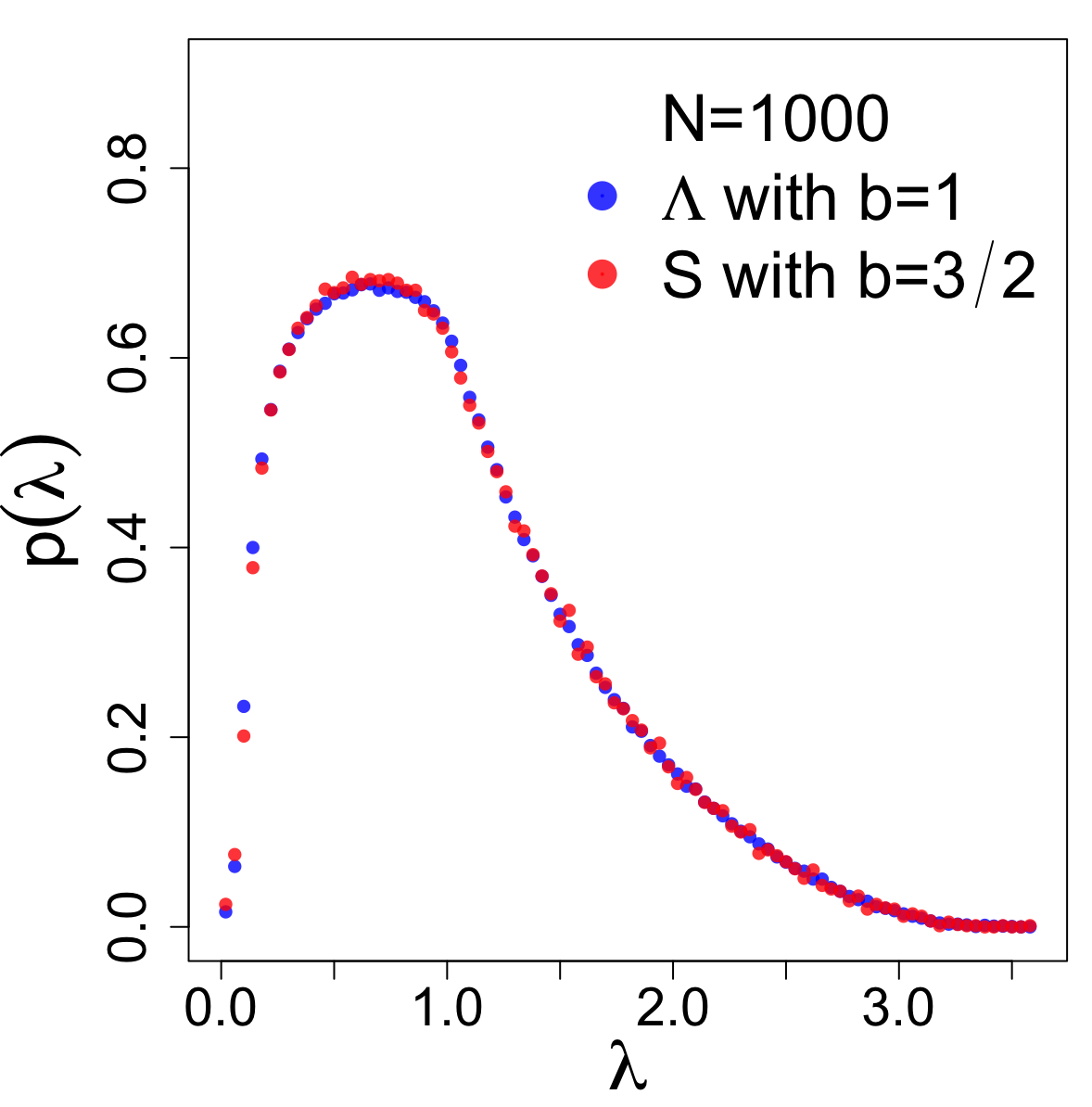}
    \includegraphics[width=0.35\textwidth]{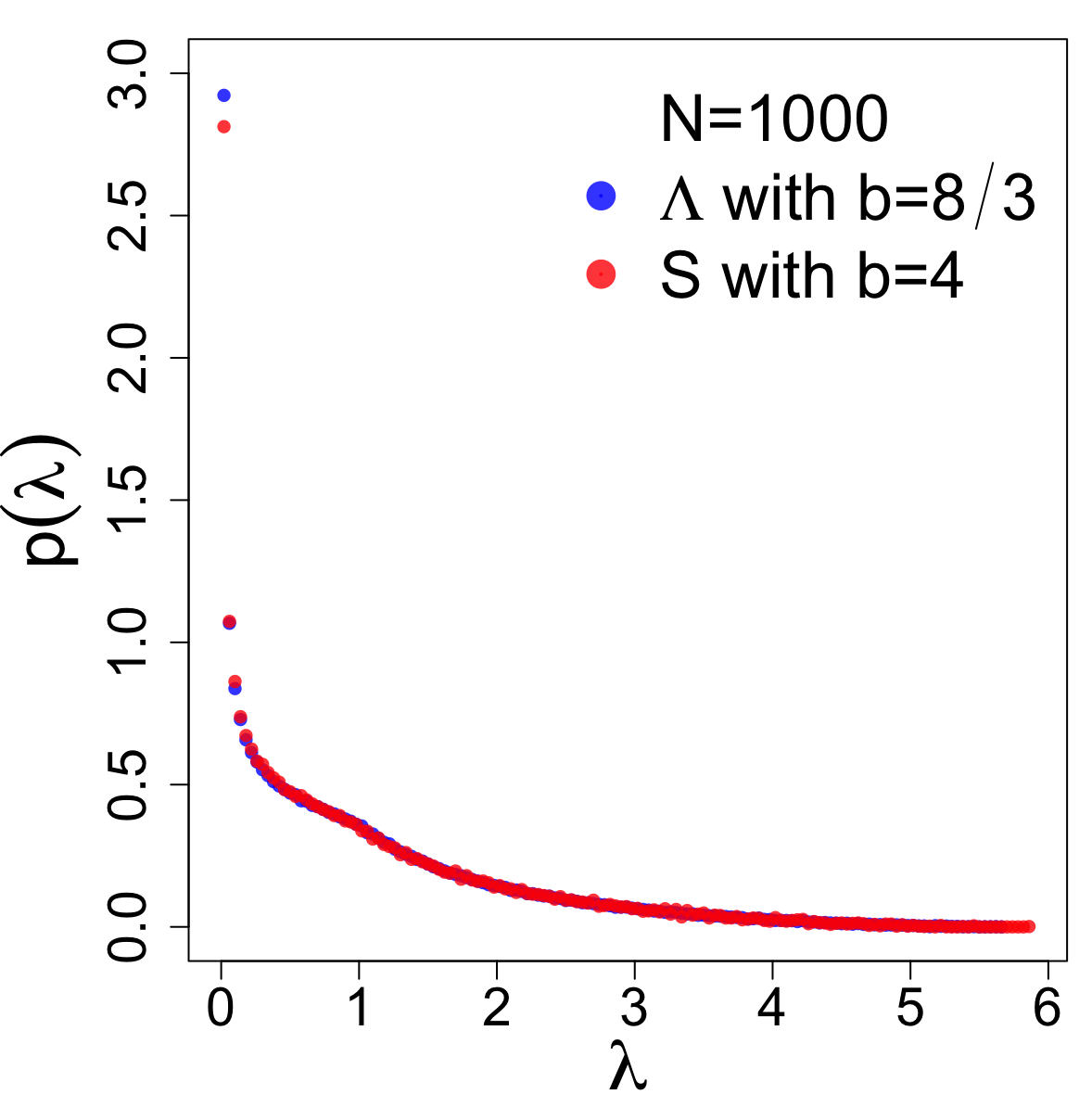}
    \includegraphics[width=0.35\textwidth]{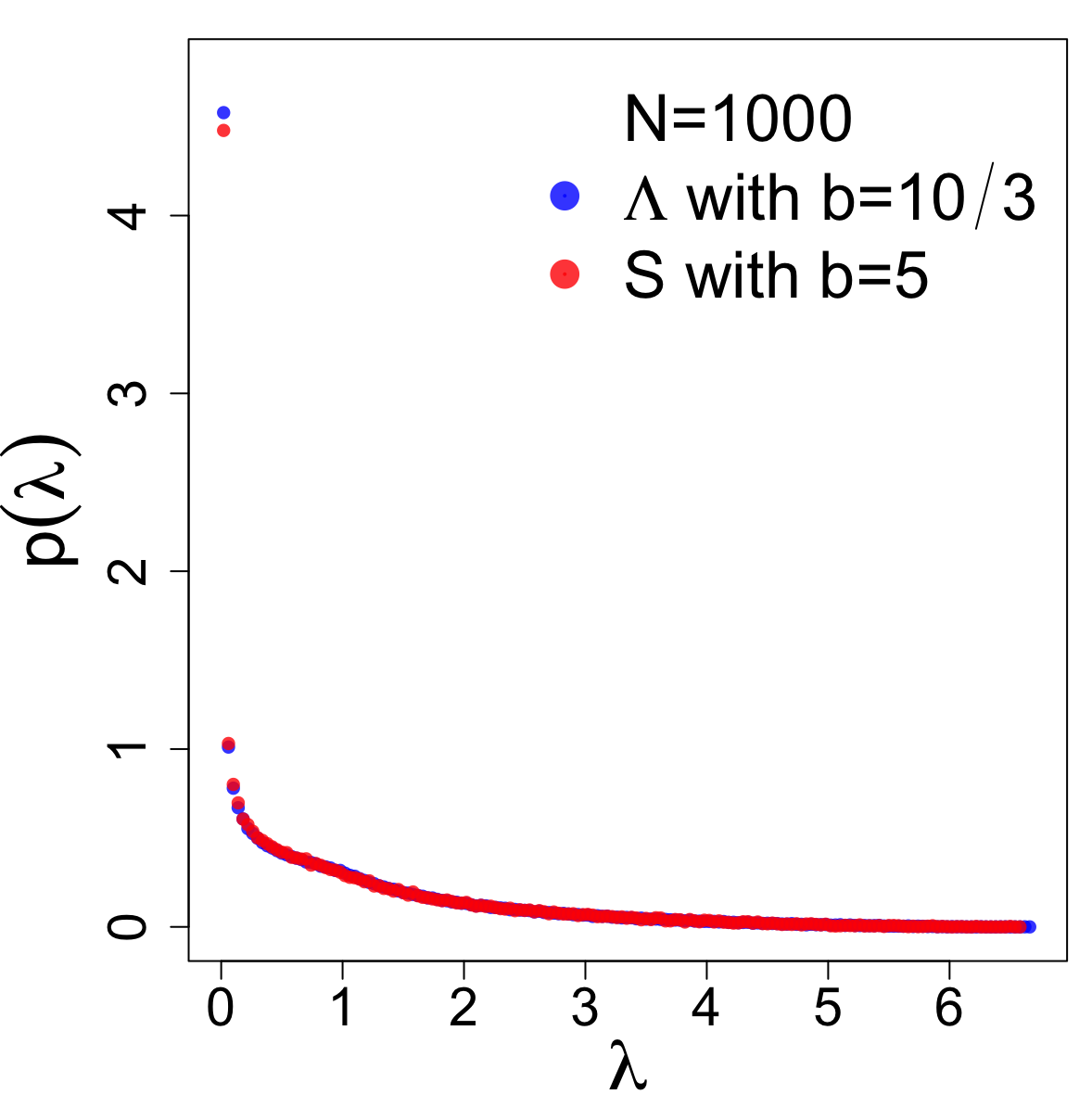}
    \caption{Comparison between the eigenvalue distributions of the scalar-model $S$ matrix and the vector-model $\Lambda$ matrix defined in Eq.~\eqref{eq:Lambda}. The $S$ matrix is diagonalized with the values $b=3/2,4,5$, while the value of $b$ for $\Lambda$ is scaled by a factor $2/3$. The distributions, represented as histograms with bin size $0.04$, are obtained by accumulating the spectra of 20 matrices each, referred to $N=1000$ atoms. Notice that the discrepancies observed in the first bin of the second panel ($0.11$) and third panel ($0.10$) are smaller than the corresponding standard errors for both the scalar-model distribution ($0.19$ for $b=4$, $0.26$ for $b=5$) and the vector-model distribution ($0.20$ for $b=8/3$, $0.20$ for $b=10/3$).}
    \label{fig:scalarvector}
\end{figure}

The above result shows that the first and second moment of the eigenvalue distributions of the scalar model matrix
\begin{equation}
    S_{ij} = \frac{\sin(k_a r_{ij})}{k_a r_{ij}} = \mathrm{sinc} \left( \sqrt{\frac{N}{b}} x_{ij} \right)
\end{equation}
coincide in the $N\to\infty$ case, up to a scaling of the parameter $b$, that must be smaller by a factor $2/3$ in the vector-model matrix (see Eq.~(22) in Ref.~\cite{Viggiano2023} for the scalar model result; see Ref.~\cite{vectorial} for considerations on the scaling factor between the two models). Moreover, in Fig.~\ref{fig:scalarvector} we numerically show that the analogy goes well beyond the second moment: in the reported cases, with $b=3/2,4,5$, the two eigenvalue distributions are practically overlapped already at $N=1000$.

\section{Evaluation of the minimum eigenvalue} 
\label{app:minimum}

\begin{figure}
    \includegraphics[width=0.48\textwidth]{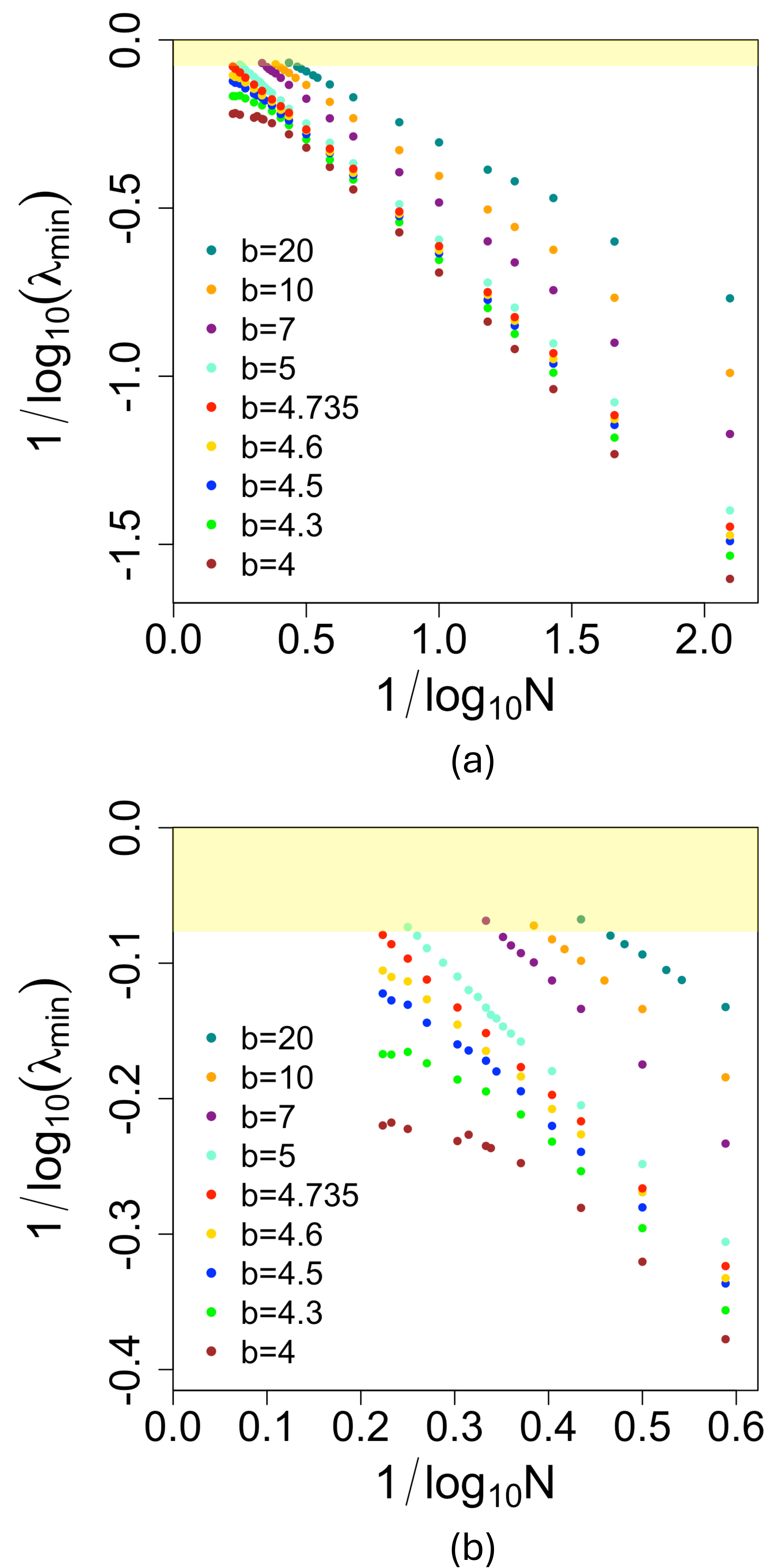}
    \caption{Plots of $1/\log_{10}(\lambda_{\mathrm{min}})$ vs $1/\log_{10} N$ obtained for different values of $b$. The horizontal yellow region corresponds to the working precision $10^{-13}$ of the diagonalization algorithm. The eigenvalues falling in this area are considered to be zero within the numerical precision.}\label{fig:minlambda}
\end{figure}

Here we further detail the protocol to obtain the minimum eigenvalue $\lambda_{\mathrm{min}}$ presented in Fig.~3 of the main text, and its scaling.
At a fixed value of $b$ we compute the minimum eigenvalue obtained from the diagonalization of $S$, and average it over several realizations of the particle positions.
The behavior of $\lambda_{\mathrm{min}}$ with the atom number $N$ is shown in Fig.~\ref{fig:minlambda}. We emphasize that all eigenvalues must be non-negative. 

For $b \geq 4.735$ we find an accurate linear dependence between $1/\log_{10}(\lambda_{\mathrm{min}})$ and $1/\log_{10} N$. It is worth noticing that such a scaling has been predicted in infinite-dimensional localization~\cite{infdloc,RRG1,RRG2}. 
Remarkably, the linear fit holds for small values of $N$ as well. This fit is valid up to the working precision $10^{-13}$ of the numerical diagonalization algorithm (yellow area in the figure). 

For $b < 4.735$ the dependence is no longer linear [see  Fig.~\ref{fig:minlambda}(b)], but rather bends towards a nonvanishing asymptotic value 
(that is, $\lambda^\infty_{\mathrm{min}} > 0$), indicating the presence of a gap.
One may try to estimate such an asymptotic value, but this would depend on some (arbitrary) nonlinear fitting function. 
Figure 3 of the main text displays the smallest (average) $\lambda_{\mathrm{min}}$ that we are able to simulate, corresponding to the case $N = 30000$. Since all curves tend to decrease monotonically with $1/\log_{10} N$, this yields a conservative estimate for $\lambda^\infty_{\mathrm{min}}$. Nonetheless, Fig.~3 of the main text displays a vertical tangent in correspondence of the critical point $b_c$.

The values of $\lambda_{\mathrm{min}} $ obtained for $b > b_c = 4.735$, numerically obtained for smaller values of $N$, cannot be distinguished from 0 within our numerical precision. Yet, above the threshold, as can be observed in Fig.~\ref{fig:minlambda}(a), it is expected to go to 0 in the $N\to\infty$ limit.

Finally, we note that, as emphasized in the text, it is by no means obvious that the fraction of condensate and the gap of the minimum eigenvalue point to the same phase transition at $b=b_c$. The simultaneous occurrence rather suggests that they are complementary manifestations of the phase transition.

\section{A perturbative estimate of the transition}
\label{app:estimate}

One can obtain a rough understanding of the nature of the transition with the following reasoning, which is perturbative in $b\ll 1$. Notice that for $b=0$ the matrix is diagonal with a degenerate spectrum $\lambda_i=1$. For small $b$, the off-diagonal matrix elements are small, with typical value
\begin{equation}
    \label{eq:AppS}
    S_{ij}\simeq \frac{(-1)^{\phi_{ij}}}{\sqrt{N/b} \,\ell_{ij} },
\end{equation}
where $\ell_{ij}=\|{\vb*{x}_i - \vb*{x}_j}\|$ is the distance between points $i,j$. We approximate the $\sin(\sqrt{N/b} \,\ell_{ij})$ with a random sign. The transition occurs when the edge of the band of eigenvalues reaches $\lambda=0$. Notice that if the matrix were not positive definite, the band would go from localized around $\lambda=1$ to being a semicircle of size $\sqrt{N}$ centered around 1. 

The eigenvalue density is expected to be
\begin{equation}
    p(\lambda)=\frac{1}{\pi N}\sum_i\Im{\bra{i}\frac{1}{S-\lambda-i 0^+}\ket{i}}.
\end{equation}
The value $p(0)$ will be non-zero only if there are delocalized states at $\lambda=0$, while all the unperturbed localized states are at energy $\lambda^{(0)}_i=1$. Working in the localization perturbation theory \cite{forward1,qrem}, the condition for this to happen is that the amplitude to get from any localized state $i$ to any other $j$ becomes of $O(1)$. In perturbation theory this is proportional to
\begin{equation}
    A_{i\to j}=\sum_{p\in {\rm Path}(i\to j)} \prod_{(kl)\in p}\frac{S_{kl}}{\lambda-\lambda_l^{(0)}}.
\end{equation}
Using the approximation (\ref{eq:AppS}), and setting $\lambda=0$ and $\lambda_l^{(0)}=1$, we obtain
\begin{equation}
    A_{i\to j}= \sum_{p}(-1)^{\phi_{p}} \left(\frac{b^{1/2}}{\sqrt{N}}\right)^N\prod_{(kl)\in p}\frac{1}{\ell_{kl}},
\end{equation}
where $\phi_p$ is the random sign assigned to the path $p$'s amplitude.

The presence of $\sqrt{N}$ in the denominators means that any single path will contribute a vanishing fraction to this sum. The largest contribution will come from the non-repeating paths that are largest in number. In this case it is easy to see that choosing any sequence $i\to i_1\to ...\to i_{N-2}\to j$ gives $(N-2)!\sim N!$ paths. 
Now we must assume  that $\ell_{kl}$ are independent and identically distributed numbers, which leads to 
\begin{equation}
    A_{i\to j}\simeq \sum_{p=1}^{N!}(-1)^{\phi_{p}}\left(\frac{b^{1/2}}{\sqrt{N}}\right)^N\frac{1}{\bar{\ell}^N},
\end{equation}
where $\bar{\ell}$ is the geometric average of $\ell_{kl}$ which can be computed (under assumption of statistical independence) using the distribution of points in Eq.~(3) of the main text, which leads to $\bar{\ell}=\exp(1-\gamma/2)=2.037$. Hence, we find the typical value of the random amplitude to be
\begin{eqnarray}
    A_{i\to j}&\propto& \left(\frac{b^{1/2}}{\overline{\ell}}\right)^N \frac{1}{N^{N/2}} \sum_{p=1}^{N!}(-1)^{\phi_{p}}\nonumber\\
    &\sim&\left(\frac{b^{1/2}}{\bar{\ell}}\right)^N\frac{\sqrt{N!}}{N^{N/2}} \sim \left(\frac{b}{b^0_c}\right)^{N/2},
\end{eqnarray}
with
\begin{eqnarray}
    b^0_c=\overline{\ell}^2 e=11.277 .    
\end{eqnarray}
Thus, under all the above approximations, a transition can be found at a finite value of $b$ which overestimates by approximately a factor $2.4$ the one found numerically. However, reexamining the approximations, one can identify several factors (dressing of the denominators, role of large fluctuations of the amplitudes, correlations of the variables $\ell_{kl}$, etc.) which can lead to a decrease in the $b_c$ estimate. In any case, this qualitative calculation captures the origin of the transition (based on a large delocalization of the eigenstates hypothesis) and the correct scaling with $N$ (the $N$ factors of $\sqrt{N}$ which are compensated by the $\sqrt{N!}$): These properties are shared with the exact solution.

\section{Finite-size corrections for the condensate fraction}
\label{app:condensate}

\begin{figure}
\includegraphics[width=0.48\textwidth]{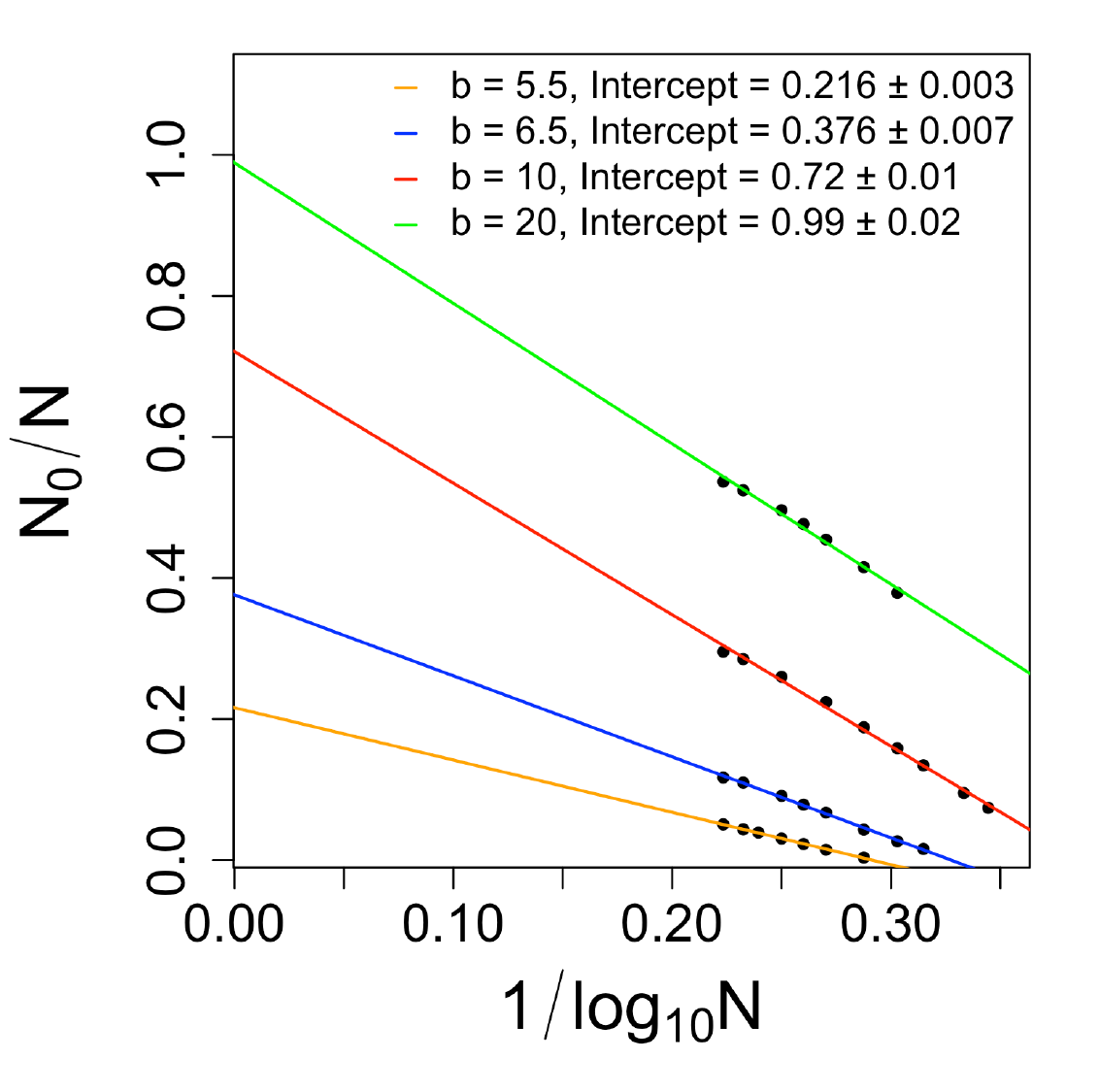}
\caption{Fraction $N_0/N$ of eigenvalues below $10^{-13}$ as a function of $1/\log_{10} N$ for $b =5.5, 6.5, 10, 20$. The intercept of the linear fits with the vertical axis corresponds to the asymptotic fraction of eigenvalues in the condensate as $N \to +\infty$. As can be seen, the intercept increases with $b$, approaching unity for $b =20$. 
}
\label{fig:esempi_picco}
\end{figure}
 
We explain here how to deal with finite-size corrections when deriving the fraction of subradiant vanishing eigenvalues for $b > b_{c}$, Fig.~4 of the main text. At a fixed value of $b$ and $N$, we count the number $N_0$ of eigenvalues below $10^{-13}$, our machine numerical precision, and compare it to the total number of eigenvalues $N$. This procedure is repeated for at least 10 realizations of the matrix $S$.
We then vary $N$. A few examples are shown in Fig.~\ref{fig:esempi_picco}, where we plot $N_0/N$ vs $(\log_{10} N)^{-1}$, obtaining a linear fit. 
This allows us to extrapolate the fitting line back to the vertical axis ($N \to \infty$). This is the value plotted in Fig.~4 of the main text.

\end{document}